\documentclass[prl,aps,a4paper,superscriptaddress,twocolumn,nofootinbib]{revtex4-2}
\usepackage{graphicx}
\usepackage{color}
\usepackage{dcolumn}
\usepackage{bm}
\usepackage{slashed}
\usepackage{amsmath}
\usepackage{latexsym}
\usepackage{amssymb}
\usepackage{mathrsfs}
\usepackage{amsfonts}
\usepackage{url}
\usepackage{hyperref}
\allowdisplaybreaks

\newcommand\prlsec[1]{\vspace{2mm}\noindent \emph{#1}.---}

\begin{document}
\title{Gravitational wave memory produced by cosmic background radiation}

\author{Zhoujian Cao}
\affiliation{Institute for Frontiers in Astronomy and Astrophysics, Beijing Normal University, Beijing 102206, China}
\affiliation{Department of Astronomy, Beijing Normal University, Beijing 100875, China}
\affiliation{School of Fundamental Physics and Mathematical Sciences, Hangzhou Institute for Advanced Study, UCAS, Hangzhou 310024, China}
\author{Xiaokai He}
\affiliation{School of Mathematics and Statistics, Hunan First Normal University, Changsha 410205, China}
\author{Zhi-Chao Zhao
\footnote{corresponding author}} \email[Zhi-Chao Zhao: ]{zhaozc@cau.edu.cn}
\affiliation{Department of Applied Physics, College of Science, China Agricultural University, Qinghua East Road, Beijing 100083, China}

\begin{abstract}
It is well known that energy fluxes will produce gravitational wave memory. The gravitational wave memory produced by background including cosmic microwave background (CMB), cosmic neutrino background (C$\nu$B), and gravitational wave background is investigated in this work. We construct a theory relating the gravitational wave memory strength to the energy flux of a stochastic background. We find that the resulted gravitational wave memory behaves as a constantly varying metric tensor. Such a varying metric tensor will introduce a quadrupole structure to the universe expansion. The gravitational wave memory due to the CMB is too small to be detected. But the gravitational wave memory due to the C$\nu$B and the gravitational wave background is marginally detectable. Interestingly, such detection can be used to estimate the neutrino masses and the properties of the gravitational wave background.
\end{abstract}

\maketitle

\prlsec{Introduction}
Gravitational wave (GW) memory is an outstanding prediction of general relativity. Such memory was firstly realized in \cite{Zeldovich74,Pay83,Braginsky:1986ia,braginsky1987gravitational}. The aforementioned memory is produced directly by the gravitational wave source and is consequently called linear memory. Later, Christodoulou \cite{PhysRevLett.67.1486,Fra92} found that gravitational wave itself can also produce memory. It means the source produces gravitational wave first and the wave produces memory then. Consequently such memory is called non-linear memory. Afterward many works have been paid to the non-linear memory of compact binary objects coalescence \cite{WisWil91,BlaDam92,Tho92,PhysRevD.95.084048}. The GW memory of compact binary objects coalescence may be observed in the near future \cite{CorJen12,MadCorCha14,Arzoumanian_2015,PhysRevLett.117.061102}. Recently people found that the non-linear memory is related to the soft theorem \cite{strominger2016gravitational,PhysRevD.89.064008,PhysRevD.95.125011,PhysRevLett.119.261602,BieCheYau11,BieGar15}. The infrared triangle makes the GW memory highly interesting to the fundamental physics \cite{PhysRevD.94.104063,PhysRevD.102.021502,PhysRevD.104.104010,PhysRevD.105.024025}.

In addition to GW, other radiations carrying energy fluxes can also produce GW memory \cite{BieCheYau11,2012BieriThe,BieGar15}. The GW memory produced by gamma ray \cite{PhysRevD.87.123007,10.1111.j.1365-2966.2005.09643.x,PhysRevD.104.104002,PhysRevD.70.104012,Yu_2020} and by neutrino emission \cite{2020ApJ...901..108V,Mukhopadhyay_2021} has been investigated. And more, many works about GW memory within the alternative gravity theory than general relativity have been done \cite{PhysRevD.103.104026,Hou2021Gravitational,PhysRevD.105.024025}. But all of the existing works on GW memory are about isolated sources. Since it has been shown that the passage of any kind of matter or radiation will cause GW memory \cite{strominger2016gravitational,PhysRevD.89.064008,PhysRevD.95.125011,PhysRevLett.119.261602,BieCheYau11,BieGar15}, it is valuable to study the GW memory produced by stochastic background radiation which can interweave fundamental physics, cosmology, and astrophysics. At the same time, such GW memory may provide a brand new means to detect GW memory which has not been detected yet. This new mean circumvents the low frequency difficulty which is related to the quasi-direct behavior of GW memory \cite{PhysRevLett.117.061102}. Unfortunately there is no study on the GW memory produced by stochastic background radiation till now. That is the topic of the current work. We construct a theory to describe the GW memory produced by the energy flux of cosmic radiation. Our theory predicts that the observed universe expansion admits a quadrupole structure due to such GW memory. To the best of our knowledge, this is the first work on the gravitational wave memory due to cosmic background radiation.

The most well known cosmic background radiation is the cosmic microwave background (CMB) which has been well detected \cite{1992ApJ...396L...1S,2003ApJS..148....1B,2016A&A...594A...1P,2020A&A...641A...6P}. It is interesting to ask how about the GW memory produced by CMB. Is it possible to detect the related GW memory? The cosmic neutrino background (C$\nu$B) should exist \cite{2006PhR...429..307L,2018NatCo...9.1833Z} based on the prediction of the standard cosmological model but has not been detected yet \cite{PhysRevD.105.063501}. It is also quite interesting to ask whether it is possible to use the related GW memory of C$\nu$B to detect or constrain the property of neutrino. Other than CMB and C$\nu$B, the stochastic background of GW (SGWB) should also exist \cite{Christensen_2018} but has not been detected too \cite{PhysRevD.104.022004,2022MNRAS.510.4873A}. It is valuable to relate the stochastic background of GW to the GW memory and use the GW memory to estimate or constraint the strength of the stochastic background of GW.

We find that the gravitational wave memory due to the CMB is too small to be detected, but the ones due to the C$\nu$B and the SGWB may be marginally detectable. Throughout this paper we will use units $c=G=1$.

\prlsec{Theory of gravitational wave memory due to cosmic background radiation}
\begin{figure}
    \centering
    \includegraphics[width=0.45\textwidth]{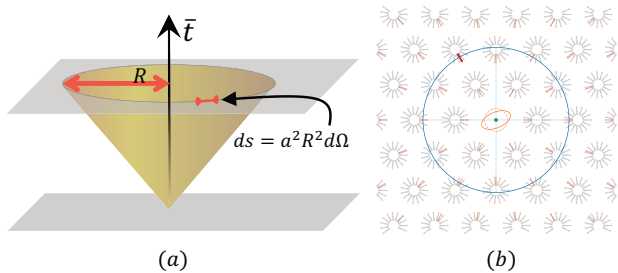}
    \caption{Diagram of the GW memory due to the cosmic background radiation. The radiation can be any type if only it carries energy flux. (a) The lightcone of the radiation relating the source and the observer. The time $\bar{t}$ corresponding to the vertical axis is the conformal time instead of the proper time. We have used $R$ to denote the comoving distance. Consequently $aR$ is the area radius of the wave front. (b) Background radiation behavior on a slice of constant cosmological time. Each source of the radiation emits energy to all directions relative to the source. The GW memory (the elliptical pattern of Universe expansion shown in the center) is produced by the sources from all directions relative to the observer. The sources corresponding to the radiation got by the observer are marked out by the blue circle.}
    \label{fig1}
\end{figure}
It is well known that the GW memory produced by GW itself can be expressed as \cite{WisWil91,BlaDam92,Tho92,PhysRevD.95.084048,Fav09a}
\begin{align}
h^{\rm m}_{ij}=\frac{4}{D_L}\int^t_{-\infty}dt'\left[\int\frac{dE^{\rm GW}}{dt'd\Omega'}\frac{n'_in'_j}{1-\textbf{n}'\cdot \textbf{N}}d\Omega'\right]^{\rm TT},
\end{align}
where TT means the transverse-traceless projection, $D_L$ is the luminosity distance between the observer and the GW source, $\textbf{n}'$ is a unit radial vector emitting from the source and $\textbf{N}$ is the unit vector pointing from the GW source to the observer. $\frac{dE^{\rm GW}}{dt'd\Omega'}$ can be related to the energy flux through
\begin{align}
\frac{dE^{\rm GW}}{dt'd\Omega'}=r^2F^\circ,
\end{align}
where $F^{\circ}$ is the energy flux radiated from the source and $r$ is the area radius of the wave front passing the observer. Using the mathematical tricks introduced in \cite{Fav09a,PhysRevD.103.043005,PhysRevD.104.064056} we can express the associated GW memory $h^{\rm m}\equiv h^{\rm m}_+-ih^{\rm m}_\times$ related to the energy flux with spin weighted -2 spherical harmonic bases ${}^{-2}Y_{lm}$ as
\begin{align}
h^{\rm m}&=\sum_{l=2}^\infty\sum_{m=-l}^{l}h^{\rm m}_{lm}[{}^{-2}Y_{lm}],\\
h^{\rm m}_{lm}&=\frac{32\pi}{D_L}\sqrt{\frac{(l-2)!}{(l+2)!}}\int^t_{-\infty}\int r^2(t')F^\circ dt'\overline{[{}^0Y_{lm}]}d\Omega', l\geq2 \label{eq8}
\end{align}
where the overline means the complex conjugate and ${}^0Y_{lm}$ is the usual spherical harmonic function (spin weight 0). The $l\geq2$ modes on the right hand side of the above equation are related to the anisotropy of the radiation. The spacetime curvature of the Friedmann-Lemaitre-Robertson-Walker metric will enhance the GW memory by a red shift factor $1+z$ \cite{Bieri_2017,PhysRevD.94.044009,2022arXiv220406981J}. Such enhancement is also true for non-memory GW \cite{1966ApJ...145..544H,1968JMP.....9..598H,2016JCAP...05..059K}. If we use $R$ to denote the comoving distance Eq.~(\ref{eq8}) becomes
\begin{align}
h^{\rm m}_{lm}&=32\pi R\sqrt{\frac{(l-2)!}{(l+2)!}}\int^t_{-\infty}\int a^2(t')F^\circ dt'\overline{[{}^0Y_{lm}]}d\Omega', l\geq2. \label{eq9}
\end{align}
where the gauge choice of the universe scale factor $a(t_0)=1$ corresponding to current observation has been chosen as usual. We emphasize that $R$ in this equation is the comoving distance instead of the luminosity distance. According to the soft theorem \cite{strominger2016gravitational,PhysRevD.89.064008,PhysRevD.95.125011,PhysRevLett.119.261602,BieCheYau11,BieGar15}, the energy flux $F^\circ$ involved in the above equation can be contributed by gravitational wave, electromagnetic wave, neutrino, and any other radiations \cite{BieCheYau11,2012BieriThe,BieGar15,PhysRevD.87.123007,10.1111.j.1365-2966.2005.09643.x,PhysRevD.104.104002,PhysRevD.70.104012,Yu_2020,2020ApJ...901..108V,Mukhopadhyay_2021}. We illustrate the physical picture of the GW memory due to the cosmic background radiation in Fig.~\ref{fig1}.

In the observer frame, the GW memory produced by the radiation from direction $(\theta,\phi)$ is
\begin{align}
&h^{\rm m}_{ij}(\theta,\phi)=\sum_{l=2}^\infty\sum_{m=-l}^{l}\nonumber\\
&\{\Re[h^m_{lm}{}^{-2}Y_{lm}(\pi-\theta,2\pi-\phi)]e^+_{ij}(\pi-\theta,2\pi-\phi)\nonumber\\
&-\Im[h^m_{lm}{}^{-2}Y_{lm}(\pi-\theta,2\pi-\phi)]e^\times_{ij}(\pi-\theta,2\pi-\phi)\},
\end{align}
where $e^{+,\times}_{ij}$ are the polarization tensors of the GW. Corresponding to a cosmic background radiation, there are similar sources in all directions. We sum all directions together
\begin{align}
h^{\rm m}_{ij}&=\int h^{\rm m}_{ij}(\theta,\phi)d\Omega.\label{eq2}
\end{align}

Straightforwardly we have
\begin{align}
&\dot{h}^{\rm m}_{ij}=\int \dot{h}^{\rm m}_{ij}(\theta,\phi)d\Omega\label{eq15}\\
&\dot{h}^{\rm m}_{ij}(\theta,\phi)=\sum_{l=2}^\infty\sum_{m=-l}^{l}\nonumber\\
&\{\Re[\dot{h}^m_{lm}{}^{-2}Y_{lm}(\pi-\theta,2\pi-\phi)]e^+_{ij}(\pi-\theta,2\pi-\phi)\nonumber\\
&-\Im[\dot{h}^m_{lm}{}^{-2}Y_{lm}(\pi-\theta,2\pi-\phi)]e^\times_{ij}(\pi-\theta,2\pi-\phi)\},\label{eq3}\\
&\dot{h}^{\rm m}_{lm}=32\pi R \sqrt{\frac{(l-2)!}{(l+2)!}}\int F^\circ \overline{[{}^0Y_{lm}]}d\Omega', l\geq2,\label{eq10}
\end{align}
where $a(t_0)=1$ has been used for the current cosmological time.

Regarding the angular integration, we express the direction dependence of the radiation as
\begin{align}
F^\circ=F_0\rho(\theta',\phi'),\\
\int\rho(\theta',\phi')d\Omega'=1.
\end{align}
Then Eq.~(\ref{eq10}) becomes
\begin{align}
\dot{h}^{\rm m}_{lm}&=32\pi F_0 R\sqrt{\frac{(l-2)!}{(l+2)!}}a_{lm}, l\geq2.\label{eq1}\\
&a_{lm}\equiv\int \rho\overline{[{}^0Y_{lm}]}d\Omega'.
\end{align}
So we can simplify Eqs.~(\ref{eq15})-(\ref{eq10}) as
\begin{align}
&\dot{h}^{\rm m}_{ij}=F_0 R\Re\left[\sum_{l=2}^\infty\sum_{m=-l}^l a_{lm}W_{lmij}\right],\\
&W_{lmij}\equiv32\pi\sqrt{\frac{(l-2)!}{(l+2)!}}\int {}^{-2}Y_{lm}(\pi-\theta,2\pi-\phi)\times\nonumber\\
&\left[e^+_{ij}(\pi-\theta,2\pi-\phi)+ie^\times_{ij}(\pi-\theta,2\pi-\phi)\right]d\Omega.
\end{align}
where $\Re$ means taking the real part.
\begin{figure}
    \centering
    \includegraphics[width=0.45\textwidth]{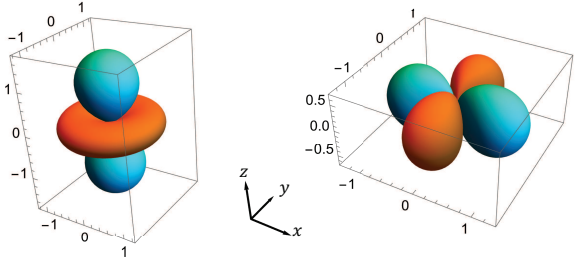}
    \caption{The quadrupole pattern for $W_{20ij}$ (left panel) and $\Re[W_{22ij}]$ (right panel). The cyan part means expansion and the orange part means contraction.}
    \label{fig2}
\end{figure}

Since $e^+_{ij}+ie^\times_{ij}$ can be expanded by harmonics $\overline{[{}^{-2}Y_{2m}]}$, all $W_{lmij}$ vanish except $W_{2mij}$. Consequently we have
\begin{align}
&\dot{h}^{\rm m}_{ij}=F_0 R\Re\left[\sum_{m=-2}^2 a_{2m}W_{2mij}\right].\label{eq6}
\end{align}
From Eq.~(\ref{eq6}) we can see only the quadrupole ($l=2$) anisotropic part of the background radiation produces GW memory. Since $W_{2mij}$s are complex matrixes, there are 9 independent bases. Except $W_{20ij}$, other 8 matrixes are related to one another through suitable rotation
\begin{align}
\Re[W_{2-2ij}]&\xrightarrow[]{\phi\rightarrow\frac{\pi}{4}+\phi}\Im[W_{2-2ij}],\\
\Re[W_{2-1ij}]&\xrightarrow[]{\phi\rightarrow\frac{\pi}{2}+\phi}\Im[W_{2-1ij}],\\
\Re[W_{21ij}]&\xrightarrow[]{\phi\rightarrow\frac{\pi}{2}-\phi}\Im[W_{2-1ij}],\\
\Re[W_{22ij}]&\xrightarrow[]{\phi\rightarrow\frac{\pi}{4}-\phi}\Im[W_{22ij}],\\
\Re[W_{2-2ij}]&=\Re[W_{22ij}],\\
\Re[W_{2-1ij}]&\xrightarrow[]{\phi\rightarrow\pi-\phi}\Re[W_{21ij}],\\
\Im[W_{2-2ij}]&\xrightarrow[]{x\rightarrow z,z\rightarrow-x}\Im[W_{2-1ij}],
\end{align}
where $\Im$ means taking imaginary part. So the 8 matrices $W_{2mij},m\neq0$ result in the same quadrupole pattern. In Fig.~\ref{fig2} we show the quadrupole pattern for $W_{20ij}$ and $\Re[W_{22ij}]$. These pattern character can be used to extract GW memory signal from kinds of experiment data.

\prlsec{Gravitational wave memory due to CMB}
According to the Stephan-Boltzmann law, the radiation flux of CMB is about $F_0\approx10^{-42}{\rm s}^{-2}$ \cite{2016A&A...594A...1P,2020A&A...641A...6P}. Since the photons decouple from the baryons at about $10^5$ years after big bang. We can estimate that the CMB source is about $R\sim 10^{17}{\rm s}$ away from us. The anisotropy modes $a_{2m}$ of CMB are about $10^{-5}$ \cite{2016A&A...594A...1P,2020A&A...641A...6P}. According to (\ref{eq6}) the strength of the gravitational wave memory due to CMB is about
\begin{align}
&\dot{h}^{\rm m_{CMB}}_{ij}\approx10^{-35}{\rm s}^{-1}.
\end{align}
Compared to the cosmic expansion rate $10^{-17}{\rm s}^{-1}$ the above correction is too tiny to be detected.

\prlsec{Gravitational wave memory due to C$\nu$B}
The massive neutrinos of the C$\nu$B are fundamental ingredients of the radiation-dominated early universe and are important non-relativistic probes of the large-scale structure formation in the late universe. We expect that the relic neutrinos have become non-relativistic at present. So the C$\nu$B abundance $\Omega_\nu$ today can be related to the sum of masses as \cite{LESGOURGUES2006307}
\begin{align}
\Omega_\nu=\frac{1}{h^2}\frac{\sum_im_{\nu_i}}{93.2{\rm eV}},\label{eq5}
\end{align}
where $h\approx0.7$ is the reduced Hubble constant. The sum of the neutrino masses has been previously restricted to the approximate range $0.06{\rm eV}\lesssim\sum_im_{\nu_i}\lesssim6{\rm eV}$ \cite{LESGOURGUES2006307}. In the following, we find that the detection of the quadrupole structure of the Universe expansion, which represents the GW memory produced by cosmic background radiation, can be used as a new probe to constrain the sum of the neutrino masses.

The flux of C$\nu$B can be estimated as
\begin{align}
F_0=\frac{3H_0^2}{8\pi}\Omega_\nu,
\end{align}
where $H_0$ means the Hubble constant. Since the neutrinos decouple from the baryons at about one second after the big bang. We can estimate that C$\nu$B source is about $R\sim 10^{17}{\rm s}$ away from us. Till now there is no detail knowledge about the anisotropy of C$\nu$B, but even if the anisotropy is small before the neutrino becomes non-relativistic, the finite mass of neutrinos amplifies the anisotropy in the low multipole moments after the non-relativistic transition. The anisotropies for a neutrino mass of 0.01 eV are amplified by more than 100 times compared to the massless case \cite{2021JCAP...06..053T}. If we assume that the anisotropy of C$\nu$B takes a value 1 for estimation, the resulting strength of the gravitational wave memory due to C$\nu$B is about
\begin{align}
&\dot{h}^{\rm m_{C\nu B}}_{ij}\approx\Omega_\nu\times10^{-18}{\rm s}^{-1}.\label{eq4}
\end{align}

Combining Eqs.~(\ref{eq5}) and (\ref{eq4}), we can use the detection of the quadrupole structure of the Universe expansion to constrain the sum of neutrino masses. Denoting the upper bound of the quadrupole structure of the Universe expansion as $\mathcal{I}$, we have
\begin{align}
\sum_im_{\nu_i}<93.2\times10^{18}\times h^2\mathcal{I}{\rm eV}.\label{eq7}
\end{align}

\prlsec{Gravitational wave memory due to SGWB}
There are many types of SGWB. Here we discuss the relic one and the one associated with CBC as examples. Standard inflation theory predicts that the relic SGWB admits strength $\rho_{\rm GW}\approx10^{-15}$ above frequency $10^{-17}{\rm Hz}$ \cite{Christensen_2018}. This prediction results in $F_0\approx10^{-50}{\rm s}^{-2}$. Since the relic SGWB is produced by inflation \cite{maggiore2018gravitational,Wang2019EstimationOS}, the comoving distance of SGWB sources is about $R\sim10^{17}{\rm s}$. Consequently, the strength of the gravitational wave memory due to relic SGWB is at most $10^{-33}{\rm s}^{-1}$ even the anisotropy takes the largest value $10$ (Each of the 9 independent bases contributes one). Although this strength is three orders larger than that of GW memory due to CMB, it is still too tiny to be detected compared to the cosmic expansion rate $10^{-17}{\rm s}^{-1}$.

Deviating from standard inflation theory, the relic SGWB may be much stronger than the one mentioned above. The combination of the latest Planck observations of CMB and lensing with baryon acoustic oscillations (BAO) and big bang nucleosynthesis (BBN) measurements can be used to determine $F_0<3.8\times10^6\times\frac{3H_0^2}{8\pi}\sim10^{-42}{\rm s}^{-2}$ \cite{2016PhLB..760..823P,PhysRevX.6.011035}. If the anisotropy of the relic SGWB can be as high as $10^{-1}$ \cite{PhysRevLett.126.141303}, the strength of the gravitational wave memory due to relic SGWB can reach about
\begin{align}
&\dot{h}^{\rm m_{relic SGWB}}_{ij}\lesssim10^{-23}{\rm s}^{-1}.
\end{align}

The strength of SGWB from compact binary coalescence (CBC) can be expressed as
\begin{align}
\Omega_{\rm GW}(f)=A_{\rm ref}\left(\frac{f}{f_{\rm ref}}\right)^\frac{2}{3},
\end{align}
which reduces to
\begin{align}
F_0=\frac{9H_0^2}{16\pi}\frac{A_{\rm ref}}{f^{\frac{2}{3}}_{\rm ref}}(f^\frac{2}{3}_{\rm merg}-f^\frac{2}{3}_{\rm form}).
\end{align}
Here $f_{\rm merg}$ and $f_{\rm form}$ correspond to the GW frequency when the CBC merges and forms respectively. In addition, $f_{\rm ref}$ and $A_{\rm ref}$ are the reference frequency and the corresponding SGWB strength which can be determined by observation. Since $f_{\rm merg}\gg f_{\rm form}$ we have
\begin{align}
F_0\approx\frac{9H_0^2}{16\pi}A_{\rm ref}\left(\frac{f_{\rm merg}}{f_{\rm ref}}\right)^{\frac{2}{3}}.
\end{align}

Regarding the stellar massive CBC which admits $f_{\rm merg}\approx10^{2}{\rm Hz}$, ground based detectors have constrained $A_{\rm ref}<10^{-9}$ at $f_{\rm ref}=25{\rm Hz}$ \cite{PhysRevD.104.022004}. This constraint reduces to $F_0\lesssim10^{-44}{\rm s}^{-2}$. About the super-massive CBC which admits $f_{\rm merg}\approx10^{-3}{\rm Hz}$, pulsar timing array (PTA) has constrained $A_{\rm ref}<10^{-6}$ at $f_{\rm ref}=10^{-8}{\rm Hz}$ \cite{lentati2015european}. This constraint reduces to $F_0\lesssim10^{-39}{\rm s}^{-2}$.

CBCs are located at different distances. If the distribution is independent of distance, the sources with larger distances dominate the GW memory contribution according to Eq.~(\ref{eq9}). We have known that galaxies have formed at least 10 billion years ago \cite{2021NatAs...5..256J}. So we can estimate that $R\sim10^{17}{\rm s}$ for CBC. Like the relic SGWB, if the anisotropy of the CBC SGWB can be as high as $10^{-1}$, the strength of the gravitational wave memory due to the stellar massive CBC and the super-massive CBC SGWB may reach respectively about
\begin{align}
&\dot{h}^{\rm m_{stelarCBC SGWB}}_{ij}\approx10^{-28}{\rm s}^{-1},\\
&\dot{h}^{\rm m_{superCBC SGWB}}_{ij}\approx10^{-23}{\rm s}^{-1}.
\end{align}

\prlsec{Gravitational wave memory due to the GW foreground of binary white dwarfs}
The GW foreground of binary white dwarfs depends strongly on the Galactic-binary merger rate and on their characteristic chirp masses, but only weakly on other parameters \cite{2012ApJ...758..131N}
\begin{eqnarray}
S_h(f) \, & \simeq & \, 1.9 \times 10^{-44} (f/\mathrm{Hz})^{-7/3} \,
\mathrm{Hz}^{-1} \label{eq:shfshort} \\
&\times&
\biggl(\frac{{\mathcal{D}_{\rm char}}}{6.4\ {\rm kpc}}\biggr)^{\!-2} \!
\biggl(\frac{\mathcal{R}_{\rm gal}}{0.015/{\rm yr}}\biggr)
\biggl(\frac{\mathcal{M}_{z, \mathrm{char}}}{0.35\, M_\odot}\biggr)^{5/3},
\nonumber
\end{eqnarray}
where $\mathcal{D}_\mathrm{char}$ is the characteristic distance from Earth to Galactic binaries; $\mathcal{R}_\mathrm{gal}$ is the binary merger rate in the Galaxy; and ${\cal M}_{z, {\mathrm char}}$ is the characteristic chirp mass. If we let these parameters take the expected values, we have
\begin{align}
F_0&\simeq 4.5\times 10^{-44}\times(f^\frac{2}{3}_{\rm up}-f^\frac{2}{3}_{\rm low}),
\end{align}
where $f_{\rm low}$ and $f_{\rm up}$ are the frequency range of the foreground. At most $f^\frac{2}{3}_{\rm up}-f^\frac{2}{3}_{\rm low}\sim1$ \cite{2022arXiv220306016A}. In addition, we can use the size of our galaxy to estimate $R\sim10^{11}{\rm s}$. The anisotropy of the foreground is at most $10^{-1}$, so the strength of the gravitational wave memory due to the foreground must be less than
\begin{align}
&\dot{h}^{\rm m_{BWD SGWB}}_{ij}\lesssim10^{-34}{\rm s}^{-1}.
\end{align}
It is too tiny to be detected.

\prlsec{Quadrupole structure of the Universe expansion}
The above analysis indicates that the gravitational wave memory due to cosmic background radiation introduces a quadrupole structure to the Universe expansion. Current observations suggest that this quadrupole should be less than $\mathcal{I}\sim10^{-18}{\rm s}^{-1}$ \cite{2022arXiv220512692D}. Plugging this constrain in Eq.~(\ref{eq7}) we get
\begin{align}
\sum_im_{\nu_i}\lesssim46{\rm eV}.
\end{align}
This constraint is consistent with previous findings \cite{LESGOURGUES2006307}. The data of the Hubble telescope and the GAIA telescope may tighten this constraint by one or two orders \cite{2021ApJ...908L...6R}. Interesting scientific discoveries about gravitational wave memory due to cosmic background radiation may be found when the detection accuracy increases in the future.

\prlsec{Discussion}
For the first time, we investigated the GW memory produced by cosmic background radiation. A theory relating GW memory to the energy flux of the cosmic background radiation is constructed based on standard general relativity. According to our theory, such GW memory behaves as a quadrupole structure of the universe expansion. So the universe expansion related measurement can be used to detect this kind of GW memory. Not like interferometer type detectors, this kind of detection can circumvent the low frequency difficulty in the GW memory detection \cite{PhysRevLett.117.061102,2021arXiv211113883Z}.

Before our work, there are less motivations to measure the quadrupole structure of the expanding universe within the standard $\Lambda$CDM cosmology \cite{2022JCAP...03..057H}. The current measurement accuracy is about $\mathcal{I}\sim10^{-18}{\rm s}^{-1}$ \cite{2022arXiv220512692D} which is almost the same level of Hubble constant itself. Significant improvements can be achieved by using the data of the Hubble telescope, the GAIA telescope, the LAMOST telescope and their combinations.

The gravitational wave memory due to the CMB is just $10^{-35}{\rm s}^{-1}$ which is too small to be detected. But the current knowledge permits that the gravitational wave memory due to the cosmic neutrino background and the gravitational wave background can be as strong as $10^{-20}{\rm s}^{-1}$ and $10^{-23}{\rm s}^{-1}$ respectively. They are marginally detectable according to the current measurement accuracy of the quadrupole structure of the universe expansion. Our work sets an important motivation to combine the data of the Hubble telescope, the GAIA telescope, the LAMOST telescope and others to measure the quadrupole structure of the universe expansion. Such measurement will not only provide direct detection of GW memory but also give important information on neutrino masses and super-massive CBC distributions.

\prlsec{Acknowledgments}
We would like to thank Paul Lasky for helpful comments on the manuscript. This work was supported in part by the National Key Research and Development Program of China Grant No. 2021YFC2203001 and in part by the NSFC (No.~11920101003, No.~12021003 and No.~12005016). Z. Cao was supported by ``the Interdiscipline Research Funds of Beijing Normal University" and CAS Project for Young Scientists in Basic Research YSBR-006. X. He is supported by the Key Project of Education Department of Hunan Province (No.21A0576).
\bibliographystyle{bibstystysty}
\bibliography{draft}

\end{document}